# Practical quantum key distribution over a 48-km optical fiber network

Richard J. Hughes, George L. Morgan and C. Glen Peterson

Physics Division
Los Alamos National Laboratory,
Los Alamos, NM 87545

## ABSTRACT

The secure distribution of the secret random bit sequences known as "key" material, is an essential precursor to their use for the encryption and decryption of confidential communications. Quantum cryptography is a new technique for secure key distribution with single-photon transmissions: Heisenberg's uncertainty principle ensures that an adversary can neither successfully tap the key transmissions, nor evade detection (eavesdropping raises the key error rate above a threshold value). We have developed experimental quantum cryptography systems based on the transmission of non-orthogonal photon states to generate shared key material over multi-kilometer optical fiber paths and over line-of-sight links. In both cases, key material is built up using the transmission of a single-photon per bit of an initial secret random sequence. A quantum-mechanically random subset of this sequence is identified, becoming the key material after a data reconciliation stage with the sender. Here we report the most recent results of our optical fiber experiment in which we have performed quantum key distribution over a 48-km optical fiber network at Los Alamos using photon interference states with the B92 and BB84 quantum key distribution protocols.

## 1. INTRODUCTION

Two of the main goals of cryptography are the encryption of messages to render them unintelligible to third parties and their authentication to certify that they have not been modified. These goals can be accomplished if the sender ("Alice") and recipient ("Bob") both possess a secret random bit sequence known as "key" material, which they use as a parameter in a cryptographic algorithm. It is essential that Alice and Bob acquire the key material with a high level of confidence that any third party ("Eve") does not have even partial information about the random bit sequence. If Alice and Bob communicate solely through classical messages it is impossible for them to generate a certifiably secret key owing to the possibility of passive eavesdropping. However, secure key distribution becomes possible if they use the single-photon communication technique of quantum cryptography, or more accurately, quantum key distribution (QKD).[1]

The security of QKD is based on the inviolability of the laws of quantum mechanics and provably secure (information theoretic) data handling protocols. Eve can neither "tap" the key transmissions owing to the indivisibility of quanta[2] nor copy them because of the quantum "no-cloning" theorem.[3] At a deeper level, QKD resists interception and retransmission by an eavesdropper because in quantum mechanics, in contrast to the classical world, the result of a measurement cannot be thought of as revealing a "possessed value" of a quantum state. A unique aspect of quantum cryptography is that Heisenberg's uncertainty principle ensures that if Eve attempts to intercept and measure Alice's quantum transmissions, her activities must produce an irreversible change in the quantum states ("collapse of the wavefunction") that are retransmitted to Bob. These changes will introduce an anomalously high error rate in the transmissions between Alice and Bob, allowing them to detect the attempted eavesdropping. In particular, from the observed error rate Alice and Bob can put an upper bound on any partial knowledge that an eavesdropper may have acquired by monitoring their transmissions. This bound allows the intended users to apply conventional information theoretic techniques to distill a secret error free key.

The first quantum cryptography protocol was published in 1984 and is now known as "BB84".[4] A further advance in theoretical quantum cryptography took place in 1991 when Ekert proposed[5] that Einstein-Podolsky-Rosen (EPR) "entangled" two-particle states could be used to implement a quantum cryptography protocol whose security was based on Bell's inequalities. Starting in 1989, Bennett, Brassard and collaborators demonstrated that QKD was potentially practical by constructing a working prototype system for the BB84 protocol, using polarized photons.[6] Although the propagation distance was only about 30 cm, this experiment is in several ways still the most thorough demonstration of quantum cryptography.

In 1992 Bennett published a "minimal" QKD scheme ("B92") and proposed that it could be implemented using single-photon interference with photons propagating for long distances over optical fibers.[7] Since then, several experimental groups [8, 9, 10, 11, 12] have developed optical fiber-based QKD systems. At Los Alamos we have demonstrated the feasibility of low-error rate QKD over underground optical fibers that were installed for network applications.[12] We have previously demonstrated QKD over 24 km of fiber[13] and here we report on results over an increased propagation distance of 48 km. QKD is also possible using line-of-sight transmissions in free-space,[14, 15, 16] which could be useful for key generation between a low-earth orbit satellite and a ground station.[16] We have developed a free-space QKD system for such applications and have achieved a transmission distance of 1-km at night[16] and more recently, 0.5 km in daylight.[17]

The remainder of this paper is organized as follows. In section 2 we give a concise introduction to the theory of quantum cryptography. Then, in section 3 we describe the experimental considerations underlying our implementation of quantum cryptography in optical fibers and the performance of our system. Finally, in section 4 we present some conclusions.

## 2. QUANTUM CRYPTOGRAPHY: THEORY

To understand QKD we must first move away from the conventional key distribution metaphor of Alice sending *particular* key data to Bob. Instead, we should have in mind a more symmetrical starting point, in which Alice and Bob initially generate their own, secret independent random binary number sequences, containing more bits than they need for the key material that they will ultimately share. They will perform a bit-wise comparison of these sequences of numbers to identify a shared random subset, which will become the key material, using a quantum transmissions over a quantum channel and a discussion of the results over a conventional, public channel. It is important to appreciate that Alice and Bob do not need to identify *all* of their shared numbers, or even *particular* ones, because the only requirements on the key material are that the numbers should be secret and random. For simplicity we shall first describe the minimal B92 QKD protocol[7] in terms of the preparation and measurement of single-photon polarization states.

Alice and Bob first agree through public discussion on how to implement the B92 protocol. For example, they can agree that Alice will transmit photons to Bob with either of two non-orthogonal polarizations: vertical polarization ("$V$") or +45° linear polarization, say. On the photons he receives, Bob can make either of two non-orthogonal polarization measurements, each of which is orthogonal to one of Alice's: -45° linear polarization or horizontal polarization ("$H$"), in this case. The second step of the protocol is for Alice and Bob to generate their independent, secret sequences of random binary numbers. They then proceed through their sequences bit-by-bit in synchronization, with Alice preparing a polarized photon for each of her bits according to the rules:

$$\begin{array}{l} "0" \leftrightarrow V \\ "1" \leftrightarrow +45° \end{array}, \qquad (1)$$

and sending it over the "quantum channel" to Bob. (The quantum channel is a transmission medium that isolates the quantum state from interactions with the "environment.") Bob makes a polarization measurement on each photon he receives, according to the value of his bit as given by:

$$\begin{matrix} "0" \leftrightarrow -45° \\ "1" \leftrightarrow H \end{matrix}, \qquad (2)$$

and records the result ("pass" = Y, "fail" = N). Note that Bob will never record a "pass" if his bit is different from Alice's, and that he records a "pass" on a random 50% portion of the bits that they have in common. For example, in Figure 1 we show Alice's preparations and Bob's measurements for the first four bits of a B92 QKD experiment.

| Alice's numbers | 1 | 0 | 1 | 0 |
|---|---|---|---|---|
| Alice's polarization | +45° | V | +45° | V |
| Bob's polarization | -45° | -45° | H | H |
| Bob's numbers | 0 | 0 | 1 | 1 |
| Bob's results | N | N | Y | N |

**Figure 1**. An example of B92 quantum key distribution

In this experiment we see that for the first and fourth bits Alice and Bob had different bit values, so that Bob's result is "N" in each case. However, for the second and third bits, Alice and Bob have the same bit values and the protocol is such that there is a probability of 0.5 that Bob's result is a "Y" in each case. Of course, we cannot predict in any particular experiment which one will be a "Y," but in this example the second bit was a "N" and the third bit was a "Y."

At this point, Bob knows that for each bit in his sequence where his result was "Y" his bit value is identical with Alice's. To complete the protocol Bob now sends a copy of his (Y or N) *results* for each bit to Alice over the public channel, but does not reveal the measurement that he made on each bit. This data reconciliation stage ends with Alice and Bob retaining only those bits for which Bob's result was "Y" and these bits become the raw material from which shared, secret key material is produced after further stages of discussion of the data over the public channel. (In the example of Figure 1 the third bit becomes the first bit of the shared key.) The B92 QKD procedure only identifies 50% of the bits that Alice and Bob actually have in common (in a perfect system), but this inefficiency is the price that Alice and Bob must pay for secrecy.

In a practical system there will be errors in the reconciled data arising from optical imperfections and detector noise, which must be removed before the key material can be used. Alice and Bob can remove these errors using conventional error correcting codes over their public channel, but at the expense of revealing some (parity) information about the resulting key material to Eve. Errors and information leakage will also occur if Eve performs her own measurement of Alice's states on the quantum channel and fabricates new photons to send on to Bob. To take an extreme case, if Eve measures each of Alice's photons using Alice's basis she will introduce a 25% error rate into Alice and Bob's key material, while correctly identifying 75% of Alice's bits. Of course, Alice and Bob could readily detect such a large error rate and would not then use their reconciled data for key material, but the eavesdropper could still gain some information at the expense of a proportionately smaller error rate if she only measures a fraction of Alice's photons. It is the goal of quantum cryptography for Alice and Bob to translate an observed error rate into an upper bound on Eve's knowledge of their reconciled data.[18] Such bounds have been established for eavesdropping attacks on individual bits[19] and are the subject of current research in the case of coherent attacks on multiple bits. Error correction can then be followed by a further stage of "privacy amplification" to reduce any partial knowledge acquired by Eve to an arbitrarily low level.[20] For example, Alice and Bob could choose the parities of random subsets of their error corrected data, and if these subsets are chosen correctly

Eve will be forced to have less than one bit of information about the resulting key. These additional stages are performed over the public channel.

Authentication of the public channel transmissions is necessary to avoid a "man-in-the-middle" attack, in which Eve could gain control of both the quantum and public channels, allowing her to masquerade as Bob to Alice and vice-versa. Alice and Bob would then unknowingly generate independent keys with Eve who could use these keys to read all of their subsequent encrypted communications. Alice and Bob need a short, secret authentication key to start the QKD procedure, and can replenish this key with a small portion of the QKD material generated. For authentication based on random hashing they will need $O(\log_2 n)$ secret authentication bits for every $n$-bit public transmission.[21]

So from the foregoing, we see that a QKD procedure may be broken down into the following seven stages:

1. Alice and Bob acquire a secret authentication key;
2. Alice and Bob generate independent secret sequences of random bits;
3. Alice and Bob use the quantum transmissions of a QKD protocol to compare their sequences and classical communications to identify a random subsequence of shared secret bits;
4. Alice and Bob perform an error correction procedure on the data;
5. Alice and Bob assess (from the error rate) how much knowledge Eve may have acquired;
6. Alice and Bob perform an appropriate privacy amplification procedure over the public channel;
7. Part of the resulting key material is used to replenish the authentication bits required in step 1, so that the system is ready for the next key generation session.

The inventors of QKD proposed that the key bits should be used for the encryption of communications using the unbreakable "one-time pad" method.[22] However, the key material could equally well (and more practically) be used by Alice and Bob in any other symmetric key cryptosystem. For example, they could use a short string of their key bits (a few hundred bits long) as an input "seed" to a cryptographically secure random number generator, whose output would provide many secure bits for use in subsequent encryption. Typically, there would no more effective method for Eve to attack this system than to exhaustively search all possible key strings, which would be computationally infeasible. Of the steps above, only one (step 3) involves the experimental physics issues that will be crucial to the practical feasibility of QKD. In our work we have therefore focussed our efforts on this component of QKD. A fully functional key generation system would include careful implementation of the other steps, but these (with the exception of step 5) are better understood and may be readily incorporated once step 3 has been adequately demonstrated. Step 5 relates to the physics of eavesdropping and a full treatment of this topic is beyond the scope of this paper. We will therefore limit ourselves to a few additional remarks on this subject.

In the simple form described above, the B92 protocol is vulnerable to Eve measuring Alice's photons in Bob's basis and only sending on those photons she can identify. (This "Bob's basis" attack would allow Eve to "force" a key onto Bob because Bob *only* detects photons for which he and Alice have the same bit value.) This will cause a factor of four reduction in bit rate unless Eve sends out multiple photons instead of just one. In the original B92 paper Bennett proposed a solution to this problem using an interferometric scheme[7] very similar to the one we have implemented in our fiber experiments. Alice and Bob could also protect against this type of attack by Bob having multiple detectors to detect for multi-photon pulses, or they could use the BB84 protocol which does not have this potential vulnerability.

In the BB84 protocol Alice generates two random bits for each photon she sends to Bob. The first bit determines which of two conjugate bases she will use for the transmission, either (H, V) or (+45º, -45º), say. The second random bit determines whether she sends a "0" or a "1", with (0, 1) = (H, V) in the first basis, or (0, 1) = (+45º, -45º) in the other basis. For each incoming photon Bob generates one random bit, which determines his measurement basis: either (H, V) or (+45°, -45°). He records whether or not a photon arrived and its polarization: H or V, or +45° or -45° depending on his basis choice. Bob then communicates to Alice over the public channel the *locations* in the photon transmission sequence where he detected photons (but not the polarization he found) and his choice of basis in each case. Notice that in BB84 in contrast to B92 Bob may incorrectly identify the bit value of a detected photon because he and Alice were using different bases. The protocol is completed with Alice informing Bob over the public channel to retain only those detected bits for which they used the same basis. The subsequence of bits for which Bob detected a photon *and* they used the same basis are perfectly correlated. Alice and Bob use this subsequence as their key. Because Alice and Bob only select the "correct basis" bits *after* Bob has detected Alice's photons, Eve has no possibility of performing a "Bob's basis" attack. The BB84 protocol is twice as efficient as B92 *per transmitted photon*. From the perspective of the physics involved B92 and BB84 are so similar that demonstration of one protocol indicates that the other will also be possible under the same physical conditions.

### 3. QUANTUM CRYPTOGRAPHY: EXPERIMENTAL REALIZATION IN OPTICAL FIBER

Although single-photon polarization states are a convenient way to explain QKD any two-state quantum system may be used. Single-photon states which are more suited to long propagation distances in optical fibers can be constructed interferometrically. Alice has a source of single photons that she can inject into a Mach-Zehnder interferometer in which she controls the phase, $\phi_A$, along one of the optical paths. Bob has single-photon detectors at the output ports (lower, "L", and upper "U") and controls the phase, $\phi_B$, along the other optical path.[1]

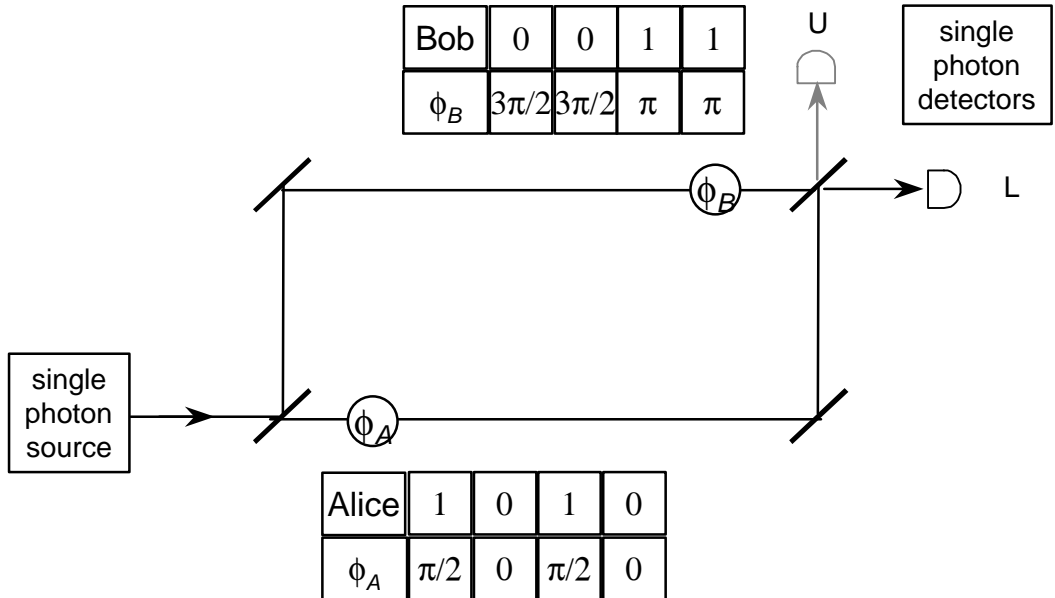

**Figure 2**. An interferometric realization of B92 quantum key distribution.

(See Figure 2 in which we have indicated the sequence of optical phases corresponding to the bit sequences in the example of section 2.) The probability that a photon injected by Alice is detected by Bob at his "L" detector

$$P_L = \cos^2\left|\frac{\phi_A - \phi_B}{2}\right| \quad , \tag{3}$$

depends on *both* paths. Thus, if Alice and Bob use the phase angles $(\phi_A, \phi_B) = (0, 3\pi/2)$ for their "0" bits (respectively) and $(\phi_A, \phi_B) = (\pi/2, \pi)$ for their "1" bits they have an exact representation of B92 when Bob records photon arrivals at his "L" detector. Each path length is analogous to one of the polarizer angles in the explanation of B92 in the previous section.

The BB84 protocol[4] can be realized with a detector in the "upper" output port, for which the single-photon detection probability is

$$P_U = \sin^2\left[\left(\frac{\phi_A - \phi_B}{2}\right)\right] \quad . \tag{4}$$

Then, Alice transmits (0, 1) in either the first basis as $\phi_A = (0, \pi)$, or the second basis as $\phi_A = (\pi/2, 3\pi/2)$, and Bob measures for photon detections at "U" or "L" with either the first basis, $\phi_B = 0$, or the second basis, $\phi_B = \pi/2$. When Alice and Bob use the same basis, Bob's "U" detector will fire to identify "1"s and his "L" detector will fire to identify "0"s.

For long transmission distances optical fibers are designed to be used at the low-attenuation wavelengths of 1.3-μm or 1.55 μm. Several groups have shown that germanium (Ge) and indium-gallium-arsenide (InGaAs) avalanche photodiodes (APDs) can detect single photons at 1.3 μm if they are first cooled to reduce noise, and operated in Geiger mode.[23] We characterized the performance of several (Fujitsu) APDs (both Ge and InGaAs) for single-photon detection at the 1.3-μm wavelength.[24] Several parameters are important in characterizing the detector performance: single-photon detection efficiency, η; intrinsic noise rate (dark counts), $R$; and time resolution. We measured absolute detection efficiencies, η, of 10 - 40%, (for InGaAs APDs), but noise rates, $R$, of 10-100 kHz. For example, the single-mode-fiber pigtailed InGaAs APDs that we use in our experiment have a noise rate that can be parameterized as

$$R(\eta) = 7.4\exp(9.2\eta) \text{ [kHz]} \quad . \tag{5}$$

in this regime, when operated in "gated mode". However, these detectors also have sub-nanosecond time resolutions, which allows us to cope with this high noise rate because of the low dispersion of optical fibers at 1.3 μm. Thus, if a 1.3-μm photon is injected into a fiber in a short wavepacket (300-ps, say) it will emerge from the far end without being significantly delocalized and so, if we know that the photon will be expected within a short time window (~ 1 ns) we need only consider the probability of a noise count in this short time interval. This probability is only ~ $5 \times 10^{-6}$ at the 50-kHz noise rate corresponding to 20% single-photon detection efficiency in the InGaAs devices.

For our fiber interferometer we use a design first proposed by Bennett[7] in which both paths are multiplexed onto a single fiber. Alice and Bob have identical, unequal-arm Mach-Zehnder interferometers with a "short" path and a "long" path, with one output port of Alice's interferometer coupled to one of the input ports of Bob's.

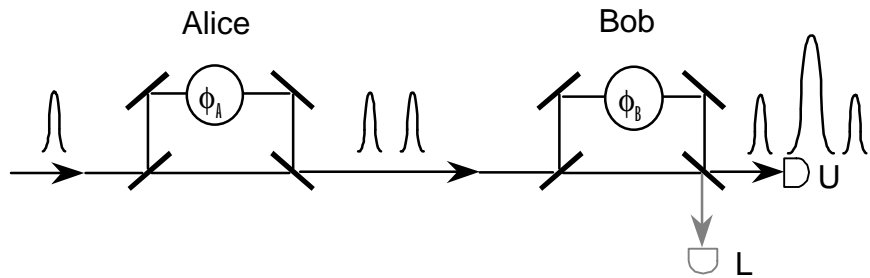

**Figure 3**. Time-multiplexed interferometer for quantum key distribution.

The difference of the light travel times between the long and short paths, $\Delta T$, is much larger than the coherence time of the light source, so there can be no interference within each small interferometer. However, interference can occur within the coupled system (see Figure 3). A photon injected into one of the input ports of Alice's interferometer therefore has a 50% probability of entering Bob's interferometer, in a wave packet that is a coherent superposition of two pieces that are separated in time by $\Delta T$, corresponding to one amplitude for it to have taken the "short" path, and a delayed amplitude which took the "long" path. On entering Bob's interferometer each component of the wave packet is again split into a "short" component and a "long" component, so that at each output port there are three "time windows" in which the photon may arrive. The first of these ("prompt") corresponds to the "short-short" propagation amplitude; which is followed after a delay of $\Delta T$ by the "central" component comprising the "short-long" and "long-short" amplitudes; and finally, after a further time $\Delta T$, the "delayed" time window corresponds to the "long-long" amplitude.

There is no interference in the "short-short" or "long-long" amplitudes, so the probability that the photon arrives in either of these time windows in either of Bob's output ports ("U" or "L") is 1/16 (we assume 50/50 beamsplitters and lossless mirrors). However, because the path-length differences in the two small interferometers are identical (to within the coherence length of the light source) interference does occur in the "central" time window between the "short-long" and "long-short" amplitudes. Indeed, because Alice and Bob can control the path length of their "long" paths with adjustable phases $\phi_A$ or $\phi_B$, respectively, the probability that the photon emerges in the "central" time window at the detector in the "U" output port in Figure 3 is

$$P_U = \frac{1}{8}\left[1 + \cos(\phi_A - \phi_B)\right] \quad . \tag{6}$$

Note that within a factor of four this expression is identical with the photon arrival probability for the simple interferometric version of B92, and that, of the two interfering paths one ("long-short") is controlled by Alice and the other ("short-long") is controlled by Bob just as in the simple interferometer of Figure 2. Thus, this time-multiplexed interferometer can be used to implement B92 QKD based on single-photon interference. (Photons arriving in the prompt and delayed time windows provide allow Alice and Bob to defeat the "Bob's basis" attack by Eve mentioned earlier.) Similarly, the probability that the photon emerges in the "central" time window at the detector in the "L" output port in Figure 3 is

$$P_L = \frac{1}{8}\left[1 - \cos(\phi_A - \phi_B)\right] \quad , \tag{7}$$

so that the BB84 protocol can be implemented by using both "U" and "L" detectors.

We have constructed an optical fiber version of this time-multiplexed interferometer in which each of Alice's and Bob's interferometers are built from two 50/50 fiber couplers. (See Figure 4.) The output fiber legs from the first coupler convey the photons to the input legs of the second coupler via a long fiber path or a short path ($\Delta T \sim 5$ ns). One of the output legs of Alice's interferometer is connected by a 48-km long underground optical fiber network path to one of the input legs of Bob's interferometer. (See Figure 5.) Photons emerge from Alice's interferometer, located in our laboratory, and are conveyed through fiber jumpers to the underground fiber network and back to Bob's interferometer, which for convenience is also located in our laboratory. The total travel time over the underground link is about 225 µs, with 22.9 dB of attenuation owing to the fiber's 0.3-dB/km attenuation and seven connections along the path. Finally, photons emerge from the output legs of Bob's interferometer into fiber pigtailed, cooled InGaAs APD detectors. (One detector in the upper output port of Figure 4 is used for B92, both output ports are used in BB84.)

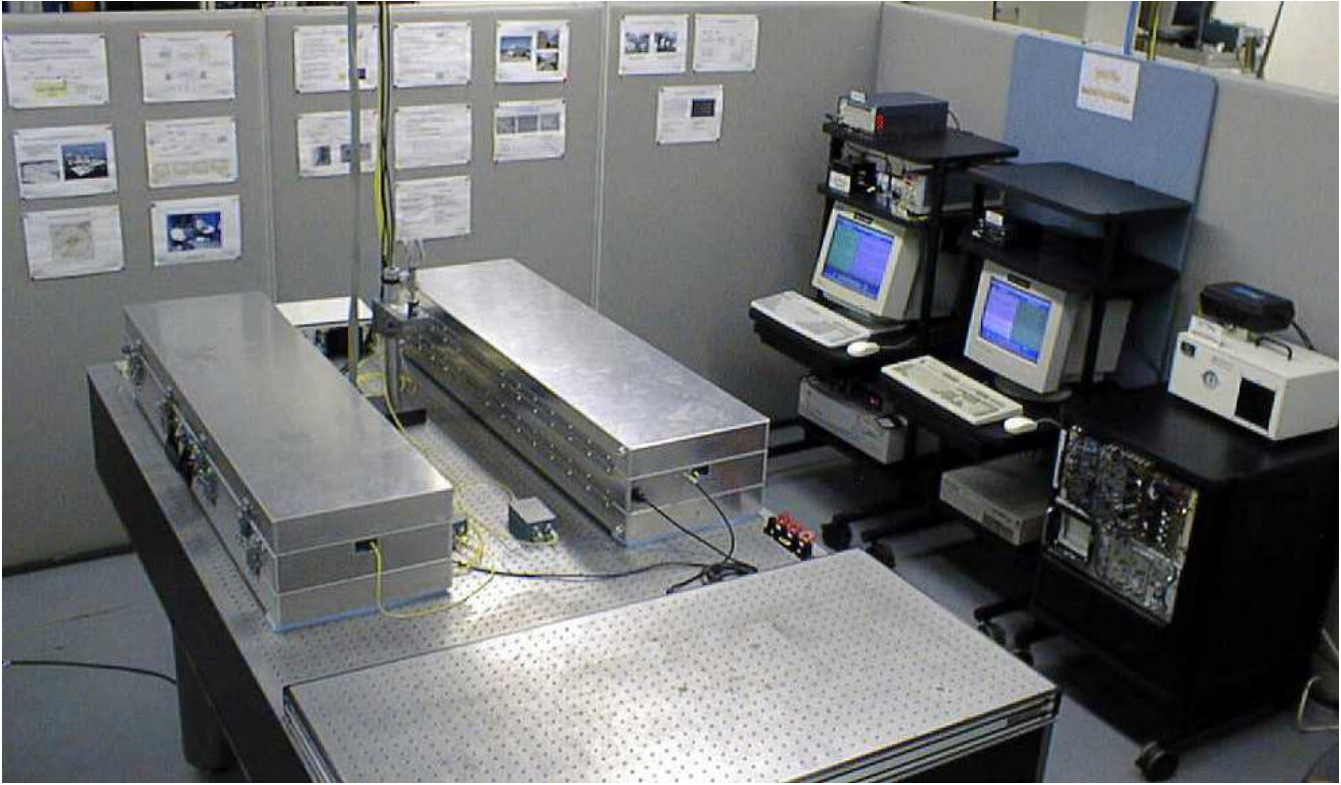

Figure 4. A view of the optical fiber quantum cryptography experiment. The two boxes in the foreground contain Alice's interferometer (box on left) and Bob's interferometer. The overhead optical fibers convey photons from Alice's control system (leftmost workstation at right rear) to Alice's interferometer, then out to the 48-km fiber network, back into Bob's interferometer and then to Bob's single photon detectors located in the refrigerator to the right of Bob's workstation (adjacent to Alice's workstation). Although physically co-located there are no direct electrical or optical connections between Alice's and Bob's systems.

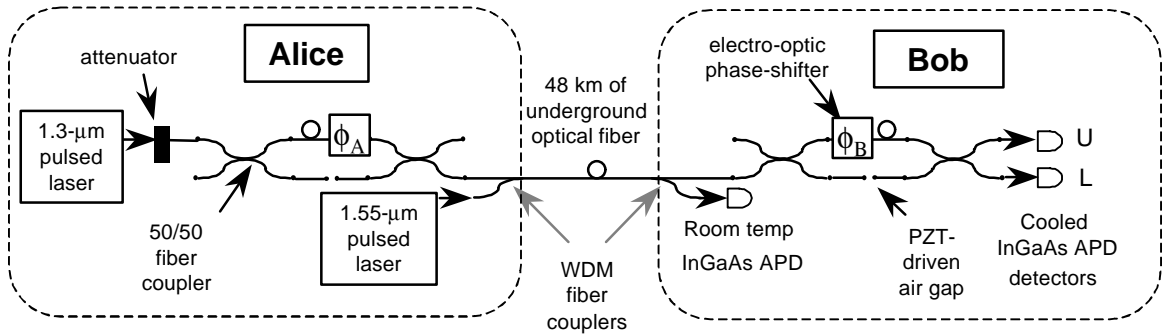

**Figure 5**. Schematic representation of the 48-km quantum cryptography experiment.

The optical path lengths in Alice's and Bob's interferometers can be adjusted in several ways. One path in each interferometer contains an adjustable air-gap for coarse adjustments, and in Bob's interferometer this air-gap is PZT-driven, which allows a stable interference to be established before key generation. A feedback signal derived from the key error rate and key bias is used to drive this air-gap and compensate for drifts during normal operations. Each interferometer also contains an electro-optic phase modulator to which short voltage pulses can be applied to achieve specific phase settings for individual photons.

A "single-photon" is generated by applying a 300-ps electrical pulse to a fiber-pigtailed 1.3-μm semiconductor laser whose output in then attenuated before coupling into the interferometer. For synchronization each "single-photon" pulse is preceded by a bright reference pulse from a 1.55-μm laser, which is introduced into the 48-km fiber path immediately after Alice's interferometer and diverted out before Bob's interferometer using wavelength-division (WDM) fiber-couplers. This bright pulse triggers a room-temperature InGaAs detector at Bob's location, which provides the "start" signal for a time-interval analyzer and triggers the pulsed-bias gate signal to his cooled single-photon detector after a delay corresponding to the single-photon emission time relative to the bright pulse emission time.

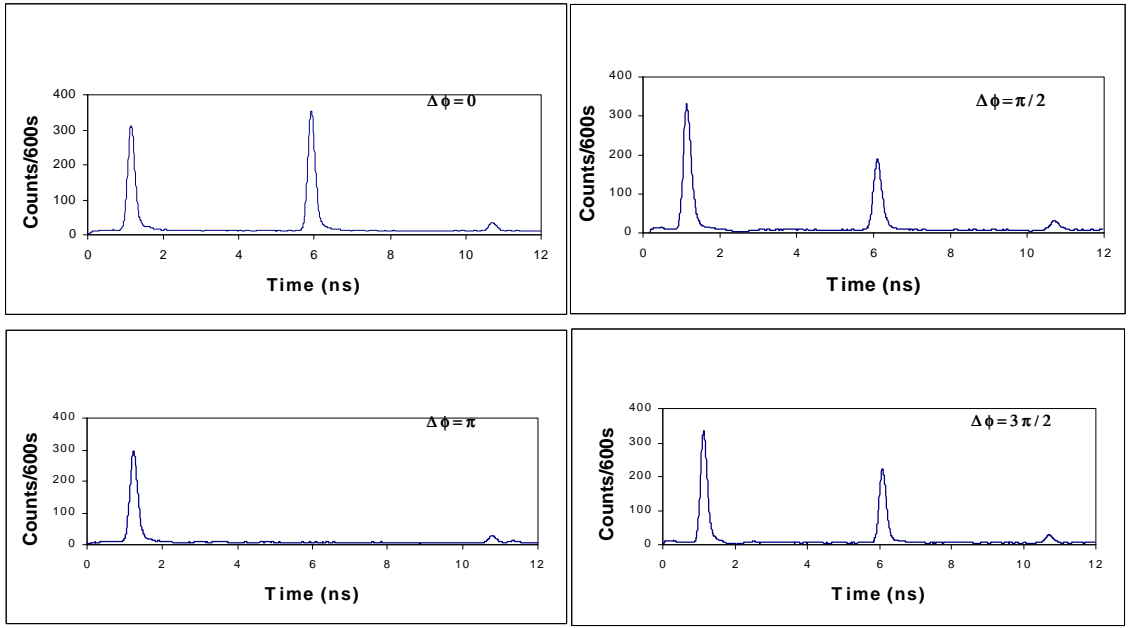

**Figure 6**. Photon time-of-arrival spectra accumulated at four phase difference values in the interferometer of Fig. 5.

Single-photon arrival times are recorded with the cold detector avalanche signal acting as the "stop" signal for the time interval analyzer. (Although Alice and Bob are located side-by-side in our laboratory there is no direct electrical connection between the sending and receiving electronics: their only links are the 48-km optical fiber "quantum channel" and the Ethernet "public channel" connection between their two independent computer control systems.) Figure 6 is an example of photon arrival time spectra at the "U" detector for four different phase differences of 0, $\pi/2$, $\pi$ and $3\pi/2$ radians, with a laser pulse rate of 100 kHz, a detection efficiency of 11%, and average central peak photon number of 0.63 leaving Alice's interferometer, assuming Poisson photon number statistics. (i.e. if Bob's interferometer was directly coupled to the output of Alice's interferometer the average photon number for light pulses arriving in the central time window at 0 radians phase difference would be 0.63.) Similar spectra are obtained using the "D" detector. Rates higher than 100 kHz led to increased detector after-pulsing noise.

Photon counts were accumulated for 600s at each phase setting. The 5-ns separation of the different paths is clearly visible, as is the 300-ps width (FWHM) of the laser pulse. The unequal height of the "short-short" (leftmost in each plot) and "long-long" (rightmost) peaks is due to attenuation in the air gaps. (This asymmetry is useful for detecting the "Bob's basis" attack by Eve.[7]) Polarization control was necessary within the interferometers in order to achieve the high visibility single-photon interference (98.99±1.24% after background subtraction) that is apparent in the central peak. In terms of B92 key generation, the 730-ps wide central peak in the $\Delta\phi = \pi/2$ graph would correspond to 10,668 bits identified as "1"s; the $\Delta\phi = 3\pi/2$ graph would correspond to 10,856 bits identified as "0"s; and the $\Delta\phi = \pi$ graph corresponds to 1,102 errors ("0"-"1" or "1"-"0").

A B92 key generation procedure starts with two independent computer control systems (Alice and Bob) generating sequences of random binary numbers by digitizing electrical noise. Each random bit is used to determine which of two voltages are applied to the electro-optic phase modulators in each interferometer for each transmitted photon. These voltages introduce the appropriate phases required in the B92 protocol. A photon pulse and a precursor bright timing pulse (which conveys no key information) are sent through the optical system for each bit. For photon events on which the "U" detector triggers, Bob can assign a bit value to Alice's transmitted bit. He records these detected bits in the memory of a computer control system, indexed by the "bright pulse" clock tick. Subsequently, Bob's computer control system transmits a file of index values (but *not* the corresponding bit values) to Alice over an Ethernet link. Alice and Bob then use those detected bits as the raw bit sequences from which an error-free, secret key is distilled using further communications over the Ethernet channel. A sample of 128 detected bits containing 6 errors is shown in Figure 7.

```
A  00100000 10011100 11111110 10010111 01110110 00000001 00101000 01111010
B  00101000 10011100 11111110 10010111 01110111 00000001 00101100 01111110

A  00010011 11001100 00111101 01101000 00110111 10110011 11101010 11011100
B  00010011 11011100 00111101 01101000 00100111 10110011 01101010 11011100
```

Figure 7: Strings of bits identified by Bob (B) and Alice (A) using the B92 protocol. (Errors marked in bold type.)

The bit error rate (BER), defined as the ratio of errors to all events, in the entire data set is ~ 9.3%. (Approximately 90% of the errors are attributable to detector dark counts.) This BER would be regarded as prohibitive in any conventional communications environment, but because of the ensured secrecy of the quantum key material it is worthwhile to take extraordinary measures to deal with such high error rates. Clearly, errors must be removed before the bit strings can be used as key material. An efficient, interactive error correction procedure has been invented that can remove *all* errors from such data sets, with BERs of up to 15%.[25] However, for simplicity in our system we perform a two-dimensional block-parity error correction procedure over the Ethernet channel, which requires Alice to reveal some parity data about the bit strings. An eavesdropper could combine this information with any knowledge acquired through eavesdropping on the quantum transmissions. There are two ways of dealing with this issue. Alice and Bob could encrypt the parity information, which would require them to have more secret bits initially, or they could perform additional privacy amplification to compensate for the information revealed, which would produce a shorter key string. We perform a rudimentary privacy amplification procedure by dropping one row and one column from each matrix of data bits. A fully functional QKD system would incorporate a more sophisticated privacy amplification procedure. At this time we have not implemented a full privacy amplification stage because our focus has been on demonstrating the essential physics involved in QKD.

We have also implemented BB84 key generation with this same system. This requires a straightforward change of the computer control software, an additional random bit generator for Alice and use of the "L" detector in Bob's receiver. The BB84 protocol achieves a factor of two greater key rate per transmitted photon than B92.

Several factors make the ~10-Hz key generation rate of our QKD system considerably slower than the laser pulse rate. Firstly, the "single-photon" requirement introduces a reduction in rate because for 53% of the laser pulses no photon leaves Alice's interferometer. The attenuation losses during propagation amount to about a factor of two hundred in our experiment. The B92 QKD procedure itself has an intrinsic inefficiency of only identifying one shared bit from four initial bits. Finally, there is the detector efficiency to be included, which in our case was 11%. (Because the key-rate increases linearly with detector efficiency, but the noise-dominated BER increases exponentially with efficiency, there is an optimal detection efficiency giving the least BER for a particular detector. This optimal efficiency is about 11%.) The low key rate is adequate for the one-time pad encryption of short text messages that we have incorporated into our software control system. But, because the one-time pad method requires as many key bits as message bits the key rate would not be adequate for more lengthy transmissions. However, this key rate would be acceptable to provide session keys for use in other symmetric key cryptosystems because such keys need only be a few hundred bits in length.

## 5. SUMMARY AND CONCLUSIONS

We have demonstrated that quantum key distribution with useful key rates is feasible over long distances (48 km) of installed optical fiber in a real-world environment. This represents more than a 50% increase in propagation distance over the longest previously reported results[9] and provides strong evidence for the practical feasibility of quantum key distribution over optical fibers for distances suitable for use within a city or a campus-like environment.

In considering possible eavesdropping on a QKD system it important to distinguish between attacks that are possible with existing technology, which are limited to individual bit attacks, and potential future attacks that are limited only by the laws of physics. In particular, all current QKD experiments use approximate single-photon states that are obtained by attenuating the output of a pulsed laser so that the average photon number per pulse is less than one. Such pulses contain a Poisson distribution of photon numbers, and the low intensity is necessary to ensure that very few pulses are vulnerable to an eavesdropper using an optical beamsplitter to "tap out" a photon from pulses containing more than one photon. In our experiment 28% of the detectable laser pulses leaving Alice's interferometer contain two or more photons because we operate at an average photon number per pulse of 0.63. This is a considerably higher fraction of multi-photon pulses than in our free-space QKD system,[16, 17] where only 6% of the detectable pulses contain two or more photons (an average photon number per pulse of 0.1). Nevertheless, even with this large multi-photon pulse probability our optical fiber QKD system could be made secure against beamsplitting and other single-bit attacks by appropriate use of privacy amplification. (We have also operated our system at a reduced average photon number per pulse of 0.39, for which only 18% of the detectable pulses contain two or more photons. In this case, the key rate dropped to ~ 3.4 Hz and the BER increased to ~ 17.8%. It is conceivable that with future development of lower noise InGaAs single-photon counting modules this BER could be significantly reduced.) However, as quantum-optical technology advances an eavesdropper could use more sophisticated methods to attack such a system in the future. For example, in a so-called QND attack,[26, 18] Eve could use a quantum non-demolition measurement to identify those pulses containing two photons. She could then determine Alice's bit value on these pulses, suppress the other pulses, and using a hypothetical lossless channel transmit a new photon to Bob. Because Alice's two-photon emission rate is larger than Bob's detection rate in our system, Bob would not notice a reduction in bit rate in this type of attack. Before such attacks become possible it will be important to replace the weak laser pulse QKD source with a true single-photon light source. Several techniques are now becoming feasible for producing such states of light. We are planning to develop a down-conversion based single-photon source to replace our pulsed, attenuated laser. Demonstration of the feasibility of QKD with weak laser pulses also implies the viability of QKD with a true single-photon light source under the same experimental conditions, because of the linearity of the processes involved.

Finally, we note that the high-visibility single-photon interference observed in our 48-km fiber interferometer allows us to place bounds on various types of loss of quantum coherence. There are two quantum mechanical amplitudes that need to be considered for a photon arriving in the "central" time window at one of Bob's detectors: the "short-long" amplitude and the "long-short" one. In an ideal system there will be coherence between these amplitudes and in the {|short-long>, |long-short>} basis we may then write the arriving photon's density matrix as,

$$\rho_{coh} = \frac{1}{2}\begin{pmatrix} 1 & ie^{i(\phi_A - \phi_B)} \\ -ie^{-i(\phi_A - \phi_B)} & 1 \end{pmatrix} , \qquad (8)$$

whereas an incoherent mixture of the amplitudes would be represented by,

$$\rho_{mix} = \frac{1}{2}\begin{pmatrix} 1 & 0 \\ 0 & 1 \end{pmatrix} . \qquad (9)$$

We may then infer a (1-$\sigma$) bound on a "collapse" of the wavefunction (8) during propagation, attributable to either unknown physical processes or hypothetical dynamical collapse mechanisms, using the parametrization,

$$\rho_{coll} = (1-p)\rho_{coh} + p\rho_{mix} , \qquad (10)$$

where $p$ is a phenomenological parameter. From our visibility data we deduce the 1-$\sigma$ limit on the collapse probability of $p < 0.0225$.

Alternatively, we may infer a bound on the loss of phase coherence using the parametrization,

$$\rho_{phase} = \frac{1}{2}\begin{pmatrix} 1 & ie^{-\xi}e^{i(\phi_A - \phi_B)} \\ -ie^{-\xi}e^{-i(\phi_A - \phi_B)} & 1 \end{pmatrix} , \qquad (11)$$

where $\xi$ is a phenomenological parameter. We find a 1-$\sigma$ limit of $\xi < 0.173$.

## ACKNOWLEDGEMENTS

It is a pleasure to thank Robert Hoffman and James Sena for their assistance in providing access to the fiber network used in these experiments. Helpful discussions with C. H. Bennett, G. Brassard, T. G. Draper, J. D. Murley, M. Kruger, P. G. Kwiat, S. K. Lamoreaux, N. Lutkenhaus, M. S. Neergaard, R. Shea and E. Twyeffort are gratefully acknowledged.

## REFERENCES

1. For reviews see R. J. Hughes et al., *Contemporary Physics* **36**, 149 (1995); C. H. Bennett et al., *Scientific American* **257** no.10, 50 (1992).;
2. J. F. Clauser, *Phys Rev D* **9**, 853 (1974).
3. W. K. Wooters and W. H. Zurek, *Nature* **299**, 802 (1982).
4. C. H. Bennett and G. Brassard, *Proceedings of IEEE International Conference on Computers, Systems and Signal Processing*, Bangalore (New York, IEEE, 1984).
5. A. K. Ekert, *Phys. Rev. Lett.* **67**, 661 (1991).
6. C. H. Bennett and G. Brassard, *SIGACT NEWS* **20**, no. 4, 78 (1989); C. H. Bennett et al., *J. Crypto.* **5**, 3 (1992).
7. C. H. Bennett, *Phys. Rev. Lett.* **68**, 3121 (1992).
8. P. D. Townsend, J. G. Rarity and P. Tapster, *Elec. Lett.* **29**, 634 (1994); P. D. Townsend, *Elec. Lett.* **30**, 809 (1994); *Opt Lett.* **20**, 1695 (1995); P. D. Townsend, *Nature* **385**, 47 (1997).
9. C. Marand and P. D. Townsend, *Opt. Lett.* **20**, 1695 (1995).
10. A. Muller et al., *Europhys. Lett.* **23**, 383 (1993); A. Muller et al., *Europhys. Lett.* **33**, 335 (1996).
11. J. D. Franson and H. Ilves, *Appl. Optics* **33**, 2949 (1994).
12. R. J. Hughes et al., *Lecture Notes in Computer Science* **1109**, 329 (1996).
13. R. J. Hughes et al., *PROC SPIE* **3076**, 2 (1997).
14. B. C. Jacobs and J. D. Franson, *Opt. Lett.* **21**, 1845 (1996).
15. W. T. Buttler et al., *Phys Rev* **A57**, 2379 (1998).
16. W. T. Buttler et al., *Phys Rev Lett* **81**, 3283 (1998).
17. R. J. Hughes et al., "Practical quantum cryptography for secure free-space communications," Los Alamos report LA-UR-99-737 (1999).
18. N. Lutkenhaus, "Estimates for practical quantum cryptography," quant-ph/9806008 (1998).
19. A. K. Ekert et al., *Phys Rev A* **50**, 1047 (1994).
20. C. H. Bennett et al., *IEEE Trans. Inf. Theory* **41**, 1915 (1995).
21. M. N. Wegman and J. L. Carter, "New hash functions and their use in authentication and set equality," J. Comp. Sys. Sci., 22, 265-279 (1981).
22. G. S. Vernam, *Trans. Am. IEE* **45**, 295 (1926).
23. P. C. M. Owens et al., *Appl. Optics* **33**, 6895 (1994); A. Lacaita et al., *Appl. Optics* **33**, 6902 (1994).
24. G. L. Morgan et al., "Study of single photon detection using germanium and indium-gallium-arsenide avalanche photodiodes," Los Alamos report LA-UR-97-4375 (1997).
25. G. Brassard and L. Salvail, "Secret-key reconciliation by public discussion,"
26. H. P. Yuen, *Quant Semicl Opt* **8**, 939 (1996).